\documentclass[11pt,A4paper]{article}
 \usepackage{graphicx}
 \usepackage{amsmath}
 \usepackage{amssymb}
 \usepackage{epsfig}
 \usepackage{natbib}    % reference
 \usepackage[utf8]{inputenc}
 \usepackage{authblk}
 \usepackage{setspace}
 \usepackage{scrextend}
\newcommand\BibTeX{{\rmfamily B\kern-.05em \textsc{i\kern-.025em b}\kern-.08em
T\kern-.1667em\lower.7ex\hbox{E}\kern-.125emX}}
\usepackage[hidelinks]{hyperref}
 \usepackage{setspace} 
\usepackage[french]{babel}
\usepackage[T1]{fontenc}
\usepackage{color}

\usepackage[
    left = \flqq{},%
    right = \frqq{},%
    leftsub = \flq{},%
    rightsub = \frq{} %
]{dirtytalk}

\usepackage{lineno}
\title{Prévisions météorologiques reposant sur l'intelligence artificielle :  \\ une révolution peut en cacher une autre. }
\author{Zied Ben Bouall\`egue$^1$, Mariana Clare$^2$, Matthieu Chevallier}
\affil[1]{Centre Européen pour les Prévisions Météorologiques à Moyen Terme, Reading, Royaume-Uni.}
\affil[2]{Centre Européen pour les Prévisions Météorologiques à Moyen Terme, Bonn, Allemagne.}
\date{}

\begin{document}

\maketitle

\begin{addmargin}[3em]{3em}
\small

\textbf{Résumé :}
 L'intelligence artificielle (IA) bouleverse aujourd'hui le monde de la prévision météorologique avec l'utilisation d'algorithmes d'apprentissage profond nourris par des champs de réanalyses. Dans ce contexte, le Centre Européen pour les Prévisions Météorologiques à Moyen Terme (CEPMMT) a décidé de développer un nouveau système de prévisions resposant sur l'IA. Ces prévisions, pour le moment de type déterministe, montrent des résultats prometteurs. Toutefois, le réalisme de ce type de prévisions reposant sur l'IA est souvent questionné. Ici, nous identifions différents types de réalisme et interrogeons notamment le rapport entre réalisme structurel et prévisibilité des évênements météorologiques. Une analyse statistique de prévisions déterministes reposant sur l'IA laisse apparaitre un dilemme réalisme/performance qu'une approche probabiliste devrait aider à résoudre.
\\
\end{addmargin}

\begin{center}
 \LARGE Weather forecasting based on artificial intelligence: \\
a revolution across the board.
\end{center}
 \begin{addmargin}[3em]{3em}
\small

\textbf{Abstract:}
Artificial intelligence (AI), based on deep-learning algorithm using high-quality reanalysis datasets, is showing enormous potential for weather forecasting. In this context, the European Centre for Medium-Range Weather Forecasts (ECMWF) is developing a new forecasting system based on AI. Verification results of deterministic forecast for now are promising. However, the realism of weather forecasts based on AI is often questioned. Here, different types of realism are identified and we discuss, in particular, the relationship between structural realism and predictability of weather events. Furthermore, a statistical analysis of deterministic forecasts based on AI points to a realism/performance dilemma that a probabilistic approach should help to solve.
\\

\end{addmargin}

\say{ La révolution tranquille des modèles numériques de prévisions météorologiques\footnote{le titre original est \textit{\textquoteleft The quiet revolution of numerical weather prediction}\textquoteright}},  tel est le titre d'un article publié dans le journal britannique \textit{Nature} en 2015 [1].  Peter Bauer et collaborateurs y relatent les progrès réalisés sur un demi-siècle en matière de prévisions météorologiques; des progrès reposant sur une série d'innovations et d'améliorations des différentes composantes des systèmes de prévisions numériques du temps.  \\

Cette révolution silencieuse est aujourd’hui rattrapée par une autre révolution qui bouleverse peu ou prou tous les domaines de la recherche et au-delà : celle de l'intelligence artificielle (IA).  \\

La figure 1 permet de mesurer les progrès accomplis ces 20 dernières années par les modèles de prévisions numériques, avec l’exemple du système de prévision du Centre Européen pour les Prévisions Météorologiques à Moyen Terme (CEPMMT, \say{ Centre Européen } ci-après) [2]. L’amélioration des prévisions s’est traduite par un gain de presque un jour de prévisibilité en 10 ans. Les prévisions à une échéance de 7 jours en 2023 sont ainsi d’une qualité presque comparable à celle des prévisions à 5 jours deux décennies plus tôt en 2003.
\\

En contrepoint, nous montrons aussi sur la figure 1 les performances du système AIFS, un nouveau système de prévisions météorologiques reposant sur l’Intelligence Artificielle (IA) que nous décrivons ci-dessous. Ce développement s'inscrit dans un contexte où de nouveaux modèles de prévision météorologique reposant sur l'IA montrent des performances prometteuses [3,4].\\

\begin{figure}[h]
\begin{center}
\includegraphics[width = 0.55\textwidth,trim={0cm 1cm 0.2cm 0cm},clip]{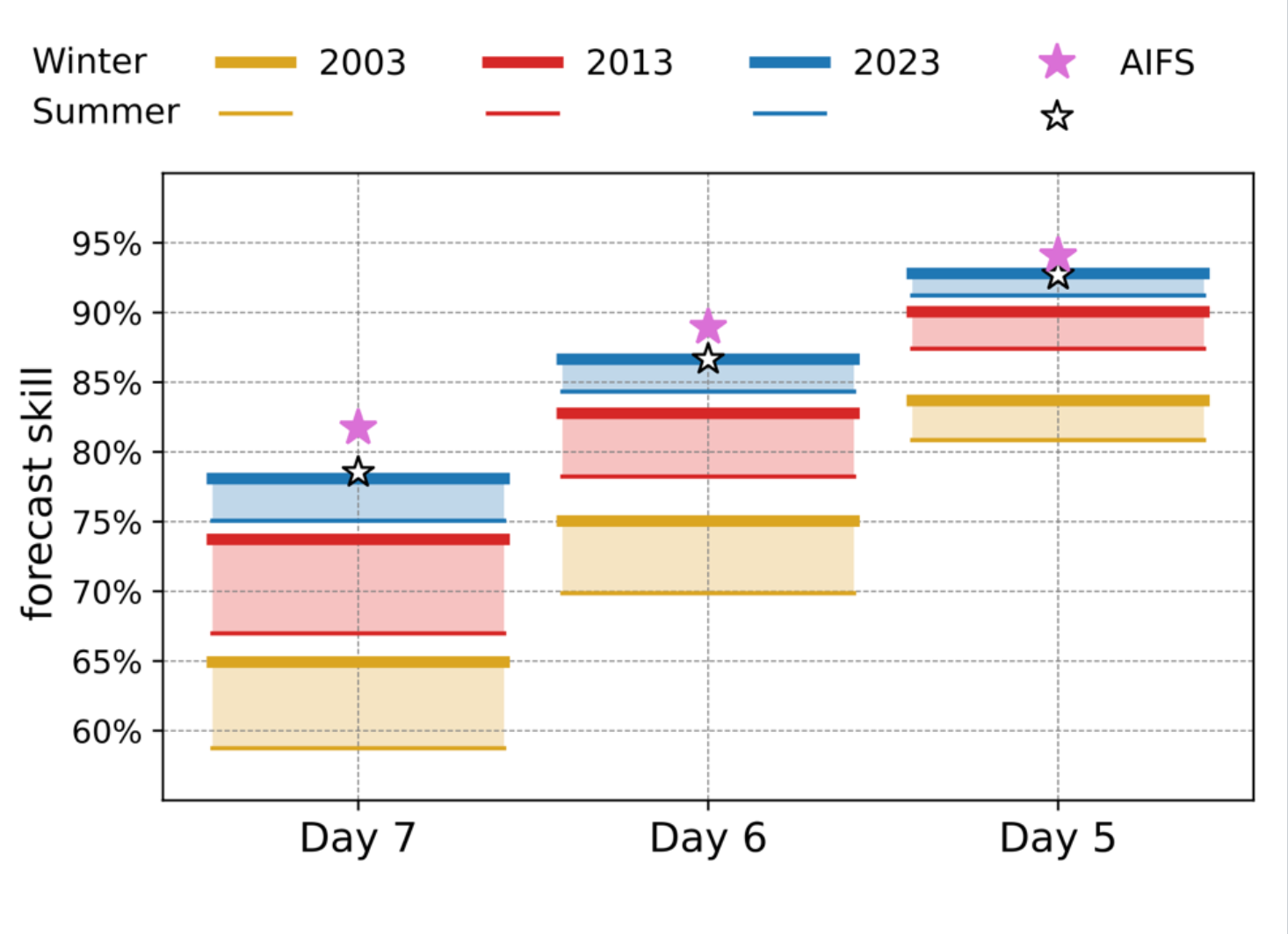}
\end{center}
\caption{
Qualité des prévisions météorologiques sur les deux dernières décennies en termes de corrélation des anomalies du géo-potentiel à 500hPa pour l'hémisphère nord, en été (Summer) et en hiver (Winter). Les résultats de AIFS pour l'année 2023 sont indiqués par une étoile.
}
\label{fig:revo}
\end{figure}

\section*{AIFS: un nouveau modèle pour le Centre Européen}

Les efforts du Centre Européen dans le domaine de l'apprentissage profond ont conduit au développement et à la mise en œuvre expérimentale d'un nouveau modèle de prévisions reposant sur l'IA.  Pendant du système numérique classique de prévisions IFS\footnote{IFS : système de prévision intégré ou \textit{Integrated Forecasting System} en anglais }, le nouveau système de prévision se dénomme AIFS (pour Artificial Intelligence/Integrated Forecasting System en anglais) [6]. \\

AIFS est entraîné principalement sur les champs de la réanalyse ERA5, la 5ème génération de réanalyse atmosphérique du Centre Européen produite par le service Changement Climatique du programme européen Copernicus [5], sur une période courant de 1978 à 2018. Les champs de l’analyse IFS opérationnelle, reposant sur la dernière version de l’IFS, sont utilisés pour un ajustement fin des coefficients du réseau de neurones sur les années 2019 et 2020. Pour le moment, AIFS repose sur une minimisation de l'erreur quadratique moyenne d'une prévision de type déterministe. Le modèle IA est entraîné à reproduire des champs globaux en partant de champs valides 6h et 12h plus tôt. Pour la dernière étape de l’entraînement, des échéances plus longues sont considérées en utilisant une approche autorégressive aussi utilisée en mode prédictif. De manière schématique, une prévision pour une échéance T est générée en partant de conditions initiales. Cette prévision au temps T sert de point de départ pour la prévision à une échéance 2T et ainsi de suite. Ce pas de temps T, dit de déploiement, est de 6h dans le cas d'AIFS. \\

Une technologie clef sous-jacente à AIFS est le transformeur sur graphe (\textit{Graph Transformer} en anglais) [7]. Le transformeur est un type de réseaux de neurones reposant sur le mécanisme d’attention qui permet un traitement efficace des relations statistiques au sein des données conduisant notamment à un gain significatif de temps d’entraînement. Quant à l’architecture de type graphique elle a l’avantage de pouvoir être utilisée sur des mailles arbitraires. Ici, l’intérêt est de permettre une représentation du globe tenant compte de sa sphéricité contrairement à une approche de grille de type « latitude-longitude » avec de nombreux points de grilles proches des pôles. Le type de maillage utilisé permet de maintenir la distance physique entre deux points de grille peu importe où l'on se situe sur le globe. Le même type de grille est utilisé dans IFS pour fournir les conditions initiales et dans les données d’entraînement ERA5. \\

La version de AIFS implémentée en janvier 2024 repose sur un maillage spatial de l'ordre de 28 km sur 13 niveaux verticaux de pression pour les variables vent, température, humidité et géo-potentiel. En surface, les sorties du modèle sont pour l'instant le cumul de précipitation total (pluie et neige), la température à 2m, le vent à 10m et la pression au niveau de la mer. Le coût approximatif d'entraînement est de 5 jours sur 64 GPUs alors que le temps de production d'une prévision déterministe à 10 jours est de l'ordre de la minute.
\\

Dans le cadre du développement de l’AIFS, un ensemble d'outils ont été regroupés sous le nom d'Anemoi\footnote{Vents en grec} : ces outils permettent l'entraînement de modèles météorologiques de manière hautement performante. Anemoi rentre dans une stratégie de mutualisation des efforts des états membres du Centre Européen autour de l'IA. Le développement de l'AIFS en bénéfice notamment [6]. Aussi, un article traitant des détails techniques de l'AIFS est en cours de préparation; les lecteurs intéressés pourront s'y référer en temps voulu.\\

\section*{AIFS et le réalisme des prévisions reposant sur l’IA }

Pour illustrer notre propos, nous prendrons comme cas d'étude la tempête Isha qui a touché les îles britanniques les 21 et 22 janvier 2024. La figure 2 compare les prévisions de IFS et AIFS à une échéance de 2 jours, c'est-à-dire initialisées le 20 janvier à 0h TU et valides le 22 janvier à 0h TU. Les champs de l'analyse opérationnelle et de la réanalyse ERA5 valides au 22 janvier y sont également montrés comme référence. \\

Des différences liées à la résolution spatiale des 2 modèles sont clairement visibles. Les champs de IFS et de l'analyse correspondante sont tous deux à une résolution d'environ 9 km. Ces champs offrent un certain niveau de détails des structures spatiales. Aussi, l'intensité maximale du vent y est proche de 30 m/s – environ 108 km/h. Les champs de AIFS et de la réanalyse ERA5 sont tous deux à une résolution d'environ 28 km et montrent moins de variabilité spatiale, ainsi que des valeurs maximales de vent plus faibles, en particulier pour la prévision AIFS qui n'excède pas 22 m/s – environ 80 km/h.  \\

Partant de cet exemple, nous pouvons distinguer trois formes de réalismes : 1) un réalisme fonctionnel nécessaire à la transmission d'une information utile à l'utilisateur, 2) un réalisme structurel qui recouvre aussi bien la notion de forme que celle d'intensité des éléments météorologiques représentés, et 3) un réalisme physique qui garantirait une cohérence spatio-temporelle et inter-variable de la prévision. Dans cette approche du réalisme des prévisions, une hiérarchie de complexité se dessine : le réalisme fonctionnel semble préexister à celui structurel qui lui-même précède, semble plus facile à atteindre qu’un réalisme physique pour les sorties de modèles d'IA.   \\

Dans l’exemple en figure 2, un réalisme fonctionnel est à l’œuvre : la tempête Isha, sa position, son creusement, sont prédits par le modèle d'IA avec un réalisme (fonctionnel) confondant. Des prévisions reposant sur l’IA peuvent en effet aujourd'hui rivaliser avec des prévisions classiques en termes de contenu informatif, même en relation avec des évènements extrêmes [4]. Cependant leur réalisme structurel semble aujourd’hui limité par une résolution spatiale encore trop grossière qui restreint leur capacité à capter de manière précise des extrêmes locaux. Par contre, le réalisme physique ne semble pas pour le moment être au rendez-vous pour les sorties de modèles IA sans contraintes physiques [8].   \\

\begin{figure}[h]
\begin{center}
\includegraphics[width = 0.49\textwidth,trim={2cm 0.cm 3cm 4.2cm},clip]{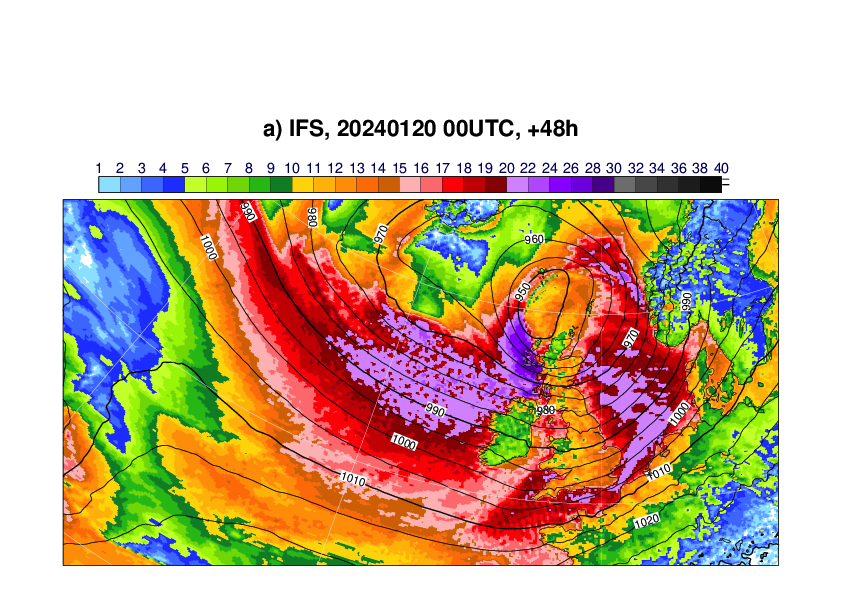}
\includegraphics[width = 0.49\textwidth,trim={2cm 0.cm 3cm 4.2cm},clip]{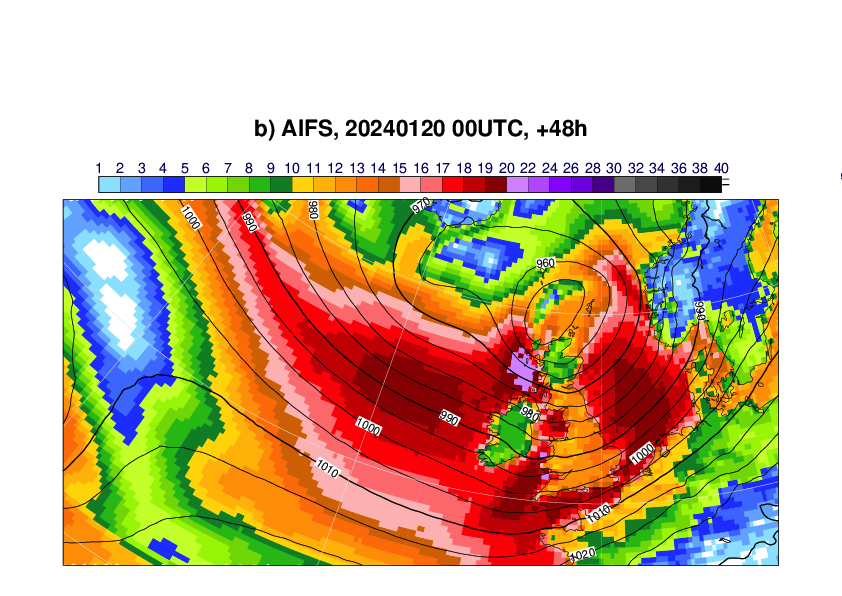}\\
\includegraphics[width = 0.49\textwidth,trim={2cm 0.cm 3cm 4.2cm},clip]{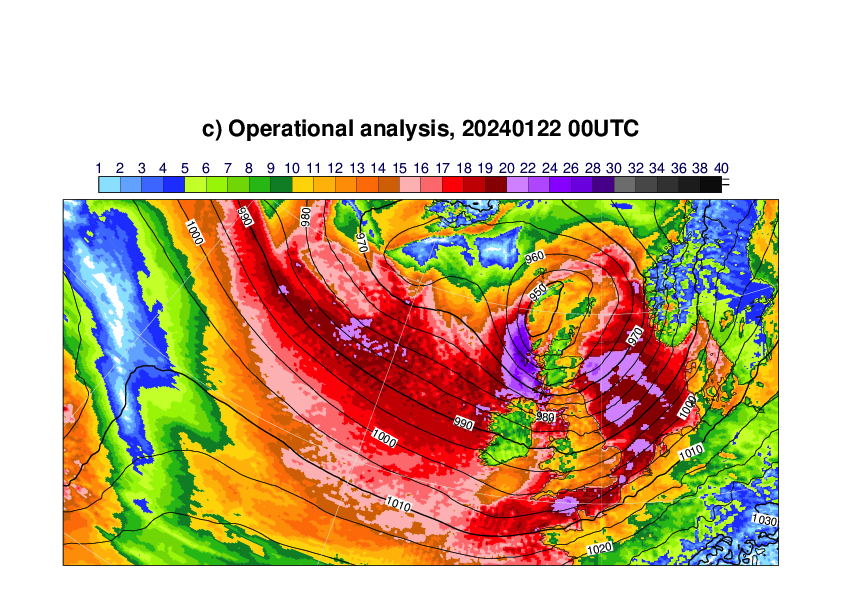}
\includegraphics[width = 0.49\textwidth,trim={2cm 0.cm 3cm 4.2cm},clip]{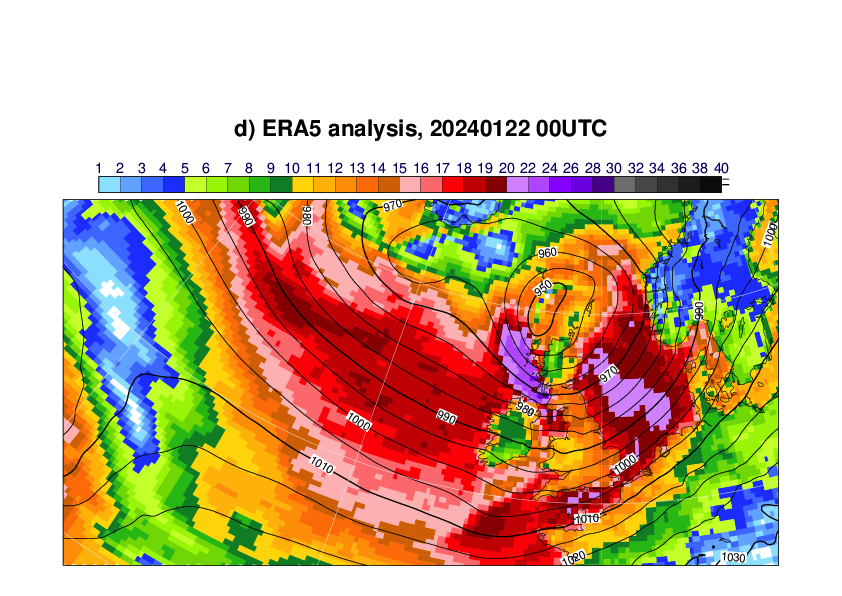}
\end{center}
\caption{
Prévisions de vents (en plages de couleur) et de la pression au niveau de la mer (en contours) à une échéance de 2 jours et champs de (ré)analyse correspondante. a) IFS, b) AIFS, c) analyse opérationnelle et d) ERA5, toutes valides pour le 22 janvier 2024.
}
\label{fig:fig}
\end{figure}

\section*{Réalisme structurel et prévisibilité}

Pour une comparaison plus quantitative, nous allons maintenant inclure 2 autres modèles d'IA dans notre analyse : GraphCast développé par Google DeepMind [9] et FuXi qui est le résultat de recherches à l’université Fudan et l’institut Qi Zhi de Shangai [10].  Les prévisions en temps réel de AIFS, GraphCast, FuXi, ainsi que d'autres modèles d'IA, sont mises en œuvre quotidiennement au Centre Européen dans une phase expérimentale, et peuvent être visualisées sur le portail de visualisation OpenCharts\footnote{https://charts.ecmwf.int}. Toutes ces prévisions sont initialisées avec l'analyse IFS opérationnelle. \\

Une approche pour évaluer le réalisme structurel d’un champ est l’analyse de sa densité spectrale de puissance. Cette analyse consiste à estimer la répartition de la puissance d'un signal suivant les fréquences qui le composent. Ainsi, la figure 3 nous montre le niveau d’énergie à des échelles allant de la méso-échelle à l’échelle synoptique. Les champs de prévisions issus de différents modèles sont analysés à 2 échéances distinctes, 2 et 6 jours. Comme référence, nous utilisons la densité spectrale de puissance des champs de l'analyse opérationnelle. Nous montrons également des résultats pour la moyenne de l'ensemble du Centre Européen discutée plus bas. \\

Pour des échéances courtes (ici 2 jours), toutes les prévisions ont un spectre similaire, proche de celui de l'analyse. Pour de plus longues échéances (ici 6 jours), nous observons moins d’énergie aux plus petites échelles, échelles auxquelles la prévisibilité est plus faible. Ceci se vérifie pour toutes les prévisions reposant ici sur l'IA mais de manière plus marquée pour FuXi.  Pour ce dernier, cette perte d’énergie à de plus longues échéances est liée au fait que FuXi repose sur différents modèles, chacun étant optimisé séparément pour différentes échéances (0-5 jours, 5-10 jours et au-delà de 10 jours).  \\

Plus l’échéance d’une prévision est longue, plus il est difficile que celle-ci soit juste et précise. En effet, l’atmosphère est un système chaotique, ce qui se traduit par une prévisibilité en général plus faible pour des horizons plus lointains. Lorsque l'état futur de l'atmosphère est particulièrement incertain, une prévision réaliste peut être fortement pénalisée par des mesures classiques de performances : une pénalité est attribuée pour avoir prévu des événements qui ne se sont pas matérialisés, à laquelle s'ajoute une pénalité pour ne pas avoir prévu ceux observés. Cette situation est bien connue des experts en vérification, qui s’y réfèrent comme étant une situation de double peine. Ce phénomène statistique conduit à encourager des prévisions floues ou peu précises, autrement dit peu réalistes tant d’un point de vue structurel que physique. De telles prévisions auront une erreur moyenne plus petite que des prévisions réalistes, en particulier lorsque l'incertitude est grande. C'est le cas par exemple de la moyenne de l'ensemble (noté ENS \textit{mean} sur la figure 3). \\

\begin{figure}[h]
\begin{center}
\includegraphics[width = 0.45\textwidth]{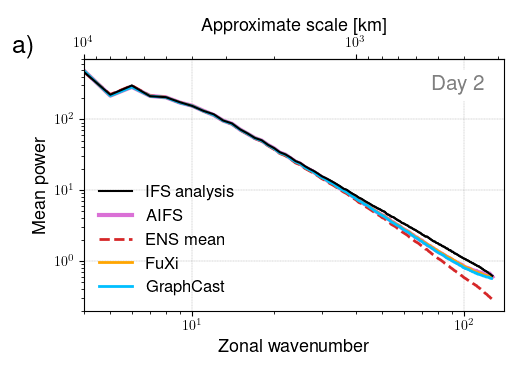}
\includegraphics[width = 0.45\textwidth]{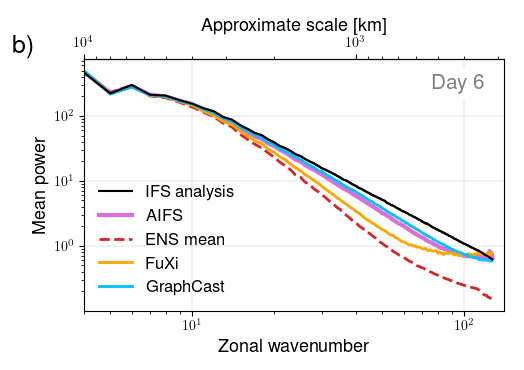}
\end{center}
\caption{Spectres de puissance pour des prévisions du géo-potentiel à 500hPa (Z500) à des échéances de : a) 2 jours et b) 6 jours comparés à ceux de l'analyse opérationnelle de l’IFS (en noir). A noter que le spectre des prévisions IFS n’est pas montré ici parce que quasi identique à celui de l’analyse. Résultats pour la période allant du 1er juin 2022 au 31 août 2022. A noter que l'échelle décroit lorsque l'abscisse croît.}
\label{fig:fig}
\end{figure}

\section*{Les limites de la prévision déterministe reposant sur l’IA }

Favoriser des prévisions floues plutôt que réalistes peut être la conséquence indésirable de la minimisation d'une fonction d’erreur, comme c'est le cas pour certains modèles d’IA. Ce risque, ou plutôt cette difficulté, se rapproche du dilemme biais-variance auquel sont confrontés les statisticiens cherchant à ajuster un modèle statistique à des données. Le modèle développé est, en quelque sorte, tiraillé entre deux injonctions contradictoires : 1) être optimal pour les données d'entraînement et 2) être généralisable à toute nouvelle situation, c'est-à-dire en mode prédictif.  \\

La figure 4 illustre ce dilemme. Les performances de 3 modèles reposant sur l'IA et la moyenne de l'ensemble y sont comparées à celle d'IFS. Nous nous attardons ici sur l'erreur quadratique moyenne comme mesure de performance et sur l'activité du modèle (défini comme l'écart type de l'anomalie de prévision) comme mesure du réalisme structurel. Le lecteur curieux ou la lectrice curieuse du formalisme mathématique de ces mesures pourra se référer à [4].  \\

Pour de courtes échéances, les modèles d’IA analysés ici parviennent à améliorer les performances par rapport à IFS sans détérioration apparente du réalisme structurel de la prévision. Cependant, à de plus longues échéances, particulièrement pour des échéances dépassant 5 jours, l’amélioration des scores RMSE s’accompagne d’une dégradation marquée de l’activité. En l’occurrence, les prévisions FuXi changent de modèle autorégressif à partir du jour 6, ce qui se traduit par une chute de l’activité qui favorise de meilleurs scores quadratiques. Sous cet angle, FuXi apparait plus proche d’une moyenne d’ensemble que d’une prévision réaliste reposant sur un modèle numérique déterministe.   \\

\begin{figure}[h]
\begin{center}
\includegraphics[width = 0.95\textwidth]{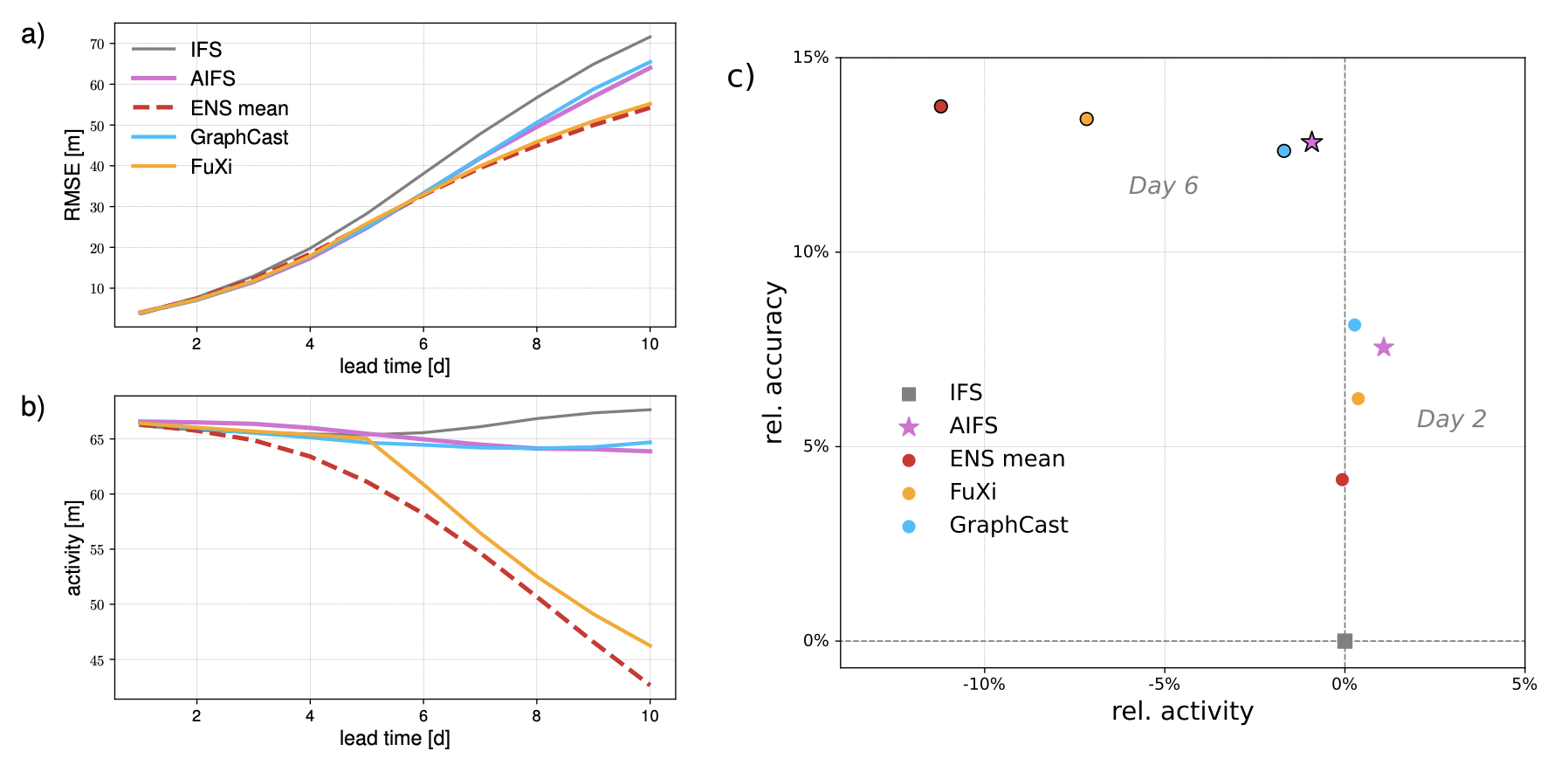}
\end{center}
\caption{
a) erreur quadratique moyenne (RMSE) et b) activité moyenne (\textit{activity}). c) Diagramme d’activité/erreur relatives (\textit{relative accuracy}) pour deux échéances : à 2 jour (Day 2) et à 6 jours (Day 6).
}
\label{fig:fig}
\end{figure}

\section*{Perspectives}
Pour sortir de ce dilemme réalisme/performances, il convient de changer de paradigme et d'adopter une approche probabiliste de la prévision météorologique. En pratique, une prévision probabiliste peut être générée à partir d'un ensemble de prévisions, une approche dite de Monte-Carlo. Chaque prévision ou membre de l'ensemble se veut une représentation réaliste de l'état de l'atmosphère pour une échéance donnée. La dispersion entre les membres de l'ensemble est une mesure de l'incertitude de la prévision. Quant à la moyenne de l'ensemble, par construction une prévision floue et peu réaliste en cas d’incertitude, celle-ci devient la prévision optimale pour minimiser l'erreur moyenne quadratique [11].\\

Les prévisions d'ensemble reposant sur des méthodes numériques classiques sont opérationnelles au Centre Européen depuis le milieu des années 90 [12].  Les incertitudes liées aux conditions initiales ainsi que les incertitudes liées à la paramétrisation des processus physiques au sein du modèle numérique y sont représentées. Avec le développement de AIFS, il est prévu de revisiter cette pratique afin de construire une prévision d'ensemble resposant sur l'IA. \\

Rappelons enfin que le modèle numérique de prévision, IFS, reste d’une importance capitale, avec sa résolution spatiale et temporelle, le nombre incomparable de variables prédites et un certain réalisme physique assuré par construction. Ceci sans compter l'assimilation de données qui fournit non seulement les conditions initiales quotidiennement mais également le jeu de données d'apprentissage pour les modèles d’IA. \\

Au moment de l'écriture de cet article, AIFS tourne en temps réel en phase expérimentale. Les sorties de AIFS sont disponibles en données ouvertes depuis le 28 février 2024 à une résolution de 28 km [13]. La mise en œuvre en opérationnel est envisagée en fonction de l’avancée des développements, notamment de la version ensembliste de AIFS.  Il est envisagé qu’un système opérationnel ensembliste sera disponible vers la fin de l’année. A ce stade, il est essentiel de continuer à analyser les performances de AIFS, de les comparer avec celle de IFS pour bien en comprendre les forces et faiblesses – ce que permet le cadre mis en place au Centre Européen avec, entre autres, l’ouverture des données de prévision AIFS. Cela permettra de développer une vision à long-terme de la complémentarité entre les deux approches.

\section*{Bibliographie}

[1] Bauer P., Thorpe A., Brunet G. The quiet revolution of numerical weather prediction. Nature 525, 47–55 (2015). doi: 10.1038/nature14956.  \\

[2] Pailleux J., Geleyn J.-F., El Khatib R., Fischer C., Hamrud M., Thépaut J.-N., Rabier F., Andersson E., Salmond D., Burridge D., Simmons A., Courtier P., 2015. Les 25 ans du système de prévision numérique du temps IFS/Arpège. La Météorologie, 89, 18-27. doi 10.4267/2042/56594\\

[3] Lguensat R., 2023 : Les nouveaux modèles de prévision météorologique basés sur l’intelligence artificielle : opportunité ou menace ?, La Météorologie, 121, 11-15.
doi:10.37053/lameteorologie-2023-0030.\\

[4] Ben Bouallègue Z, Clare MCA, Magnusson L, Gascon E, Maier-Gerber M, Janousek M, Rodwell M, Pinault F, Dramsch JS, Lang STK, Raoult B, Rabier F, Chevallier M, Sandu I, Dueben P, Chantry M, Pappenberger F. (2023). The rise of data-driven weather forecasting. Bulletin of the American Meteorological Society. doi: 10.1175/BAMS-D-23-0162.1 \\

[5] Hersbach, H., Bell, Bill, Berrisford, P. and co-authors, 2020. The ERA5 global reanalysis, Quarterly Journal of the Royal Meteorological Society, 146, 1999-2049. doi:10.1002/qj.3803. \\

[6] Lang S., Alexe M., Chantry M., Dramsch J., Pinault F., Raoult B., Ben Bouallègue Z., Clare M, Lessig C., Magnusson L., Prieto Nemesio A., 2024. AIFS: a new ECMWF forecasting system. ECMWF Newsletter number 178, 4-5. doi:10.21957/1a8466ec2f \\

[7] Lang S., Alexe M., Chantry M., Dramsch J., Pinault F., Raoult B., Ben Bouallègue Z., Clare M, Lessig C., Magnusson L., Prieto Nemesio A., 2024. First update to the AIFS. AIFS blog post accessible at www.ecmwf.int/en/about/media-centre/aifs-blog/2024/first-update-aifs\\

[8] Bonavita M. (2023). On some limitations of data-driven weather forecasting models. ArXiv preprint. doi: 10.48550/arXiv.2309.08473. \\

[9] Lam R., Sanchez-Gonzalez A., Willson M., Wirnsberger P., Fortunato M., Pritzel A., Ravuri S., Ewalds T., Alet F., Eaton-Rosen Z., Hu W., Merose A., 2023. Learning skillful medium-range global weather forecasting. Science 382, 1416-1421. doi:10.1126/science.adi2336.  \\

[10] Chen, L., Zhong, X., Zhang, F. et al., 2023, FuXi: a cascade machine learning forecasting system for 15-day global weather forecast. npj Clim Atmos Sci 6, 190. doi: 10.1038/s41612-023-00512-1.  \\

[11] Palmer T.N., Barkmeijer J., Buizza R., Klinker E., Richardson D., 2002. L’avenir de la prévision d’ensemble. La Météorologie, 36, 22-30. doi 10.4267/2042/36203.\\

[12] Molteni, F., Buizza, R., Palmer, T.N. and Petroliagis, T., 1996. The ECMWF Ensemble Prediction System: Methodology and validation. Q.J.R. Meteorol. Soc., 122: 73-119. doi:10.1002/qj.49712252905.  \\

[13] Chantry M., Alexe M., Dramsch J., Pinault F., Raoult B., Ben Bouallègue Z., Clare M, Lang S., Lessig C., Magnusson L., Prieto Nemesio A., 2024. It's rain(ing) data. AIFS blog post accessible at

www.ecmwf.int/en/about/media-centre/aifs-blog/2024/its-raining-data\\

\end{document}